\documentclass[review, 12pt]{elsarticle}

\usepackage{graphicx}
\usepackage{dcolumn}
\usepackage{bm}
\usepackage{color}
\usepackage{xcolor}
\usepackage{epstopdf}
\usepackage{gensymb}
\usepackage{ulem}
\usepackage{float}
\usepackage{xr}
\usepackage{mathrsfs}
\usepackage{amsmath}
\usepackage{amsthm}
\usepackage{enumerate}
\usepackage{soul}
\usepackage{amssymb}
\usepackage{units}
\usepackage{todonotes}
\usepackage{verbatim}
\usepackage{hyperref}
\usepackage{sidecap}
\usepackage{wrapfig}
\usepackage{caption}
\usepackage{subcaption}
\usepackage{cleveref}
\usepackage{dsfont}
\usepackage{lipsum}
\usepackage{comment}
\usepackage{bm}
\usepackage{adjustbox}

\journal{Journal of Computational Physics}
\begin{document}

\begin{frontmatter}
\title{Measurement-Efficient Variational Quantum Linear Solver for Carleman-Linearized Nonlinear Dynamics}

\author[gt]{Yunya Liu}
\address[gt]{Woodruff School of Mechanical Engineering,
             Georgia Institute of Technology, Atlanta, Georgia, USA}
\author[ut]{Pai Wang\corref{cor1}}
\ead{pai.wang@utah.edu}
\address[ut]{Department of Mechanical Engineering, University of Utah, Salt Lake City, UT, USA}
\cortext[cor1]{Corresponding author}

\begin{abstract}
We present hybrid quantum-classical pipelines for solving the Duffing equation that leverage Carleman linearization and the Variational Quantum Linear Solver (VQLS). First, we demonstrate that Carleman linearization accurately approximates the weakly nonlinear Duffing equation, with errors diminishing as the truncation order increases. Next, across IBM and Xanadu platforms, we deploy VQLS with symmetry-grouped Hadamard Test evaluations under both global and local cost formulations, compare distinct Hermitianization within a common cost framework, and benchmark hardware-efficient ansatz architectures under a fixed Hermitianization. Across block-banded test cases, each method achieves near-unity fidelity and vanishing relative residuals. These results show that topology-agnostic ansatz, optimized Hermitianization, and efficient cost formulation enable VQLS to recover quantum states proportional to classical solutions for Carleman-structured systems, providing a portable recipe for quantum-in-the-loop simulation of nonlinear dynamics.
\end{abstract}
\end{frontmatter}

\section{Introduction}
Nonlinear differential equations pervade modern engineering analysis, arising in nonlinear structural mechanics involving large deformations or material plasticity, in fluid dynamics governed by the Navier--Stokes equations~\cite{temam2024navier}, in heat transfer with temperature-dependent properties~\cite{nellis2008heat}, and in control systems with nonlinear actuator or sensor characteristics~\cite{sastry2013nonlinear}. A canonical and widely studied representative of weakly nonlinear engineering oscillators is the damped, externally forced Duffing equation\cite{kovacic2011duffing},
\begin{equation}
    \ddot z + \delta\,\dot z + \alpha\,z + \beta\,z^3 = f(t),
    \label{eq:duffing}
\end{equation}
with damping coefficient $\delta$, linear stiffness $\alpha$, cubic stiffness $\beta$, and external forcing $f(t)$~\cite{hubay2021return}. Despite its compact form, Eq.~\eqref{eq:duffing} serves as a canonical low-dimensional polynomial ordinary differential equation (ODE) for benchmarking new nonlinear-treatment methods, owing to its cubic stiffness and its well-characterized phenomenology---including amplitude-dependent frequency shifts, jump and hysteresis phenomena, and chaotic regimes.

Solving polynomial nonlinear DEs on a computer typically requires two design choices: a strategy for treating the polynomial nonlinearity, and a numerical solver for the resulting linear or algebraic system. Classical strategies for the first include local Jacobian linearization, harmonic balance, perturbation expansions such as the Lindstedt--Poincaré and multiple-scales methods, the Koopman observable lifting, and Carleman linearization~\cite{kowalski1991nonlinear, forets2017explicit, forets2021reachability, ito2023map, surana2022efficient, amini2022carleman, shi2023koopman, chermnykh2016carleman, le2012carleman}. Classical choices for the second range from time-stepping schemes such as explicit and implicit Euler~\cite{biswas2013discussion} and Runge-Kutta methods, and series-based propagators including Taylor~\cite{berry2015simulating} and Dyson~\cite{argeri2014magnus} expansions, to discretization frameworks such as the finite element method (FEM)~\cite{reddy1993introduction}.

Among nonlinearity-treatment strategies, Carleman linearization is distinguished by its global and structural nature. It provides an exact infinite-dimensional linear embedding of polynomial-in-state ODEs; any practical implementation requires truncation at a finite order $N$, which introduces an approximation error that has been shown to decay exponentially in $N$ under dissipative conditions~\cite{liu2021efficient}, with explicit finite-section error bounds available~\cite{amini2022carleman} and recent extensions to non-dissipative resonant regimes~\cite{wu2024quantum}.
For cubic polynomial ODEs, including the Duffing equation, the truncated system is block-banded, with structured sub- and super-diagonal blocks that couple adjacent moment levels. 

This structured sparsity is particularly attractive computationally. By contrast, the closely related Koopman framework yields a comparable banded representation only when the observable dictionary is restricted to monomials, in which case it reduces to Carleman linearization itself~\cite{joseph2020koopman}. Frequency-domain harmonic-balance formulations typically give nonlinear algebraic systems with a block Toeplitz-plus-Hankel structure~\cite{yan2023harmonic}; classical Jacobian linearization provides a single Jacobian-based linear system at each linearization point or Newton iteration~\cite{cordero2024two}; and perturbation methods typically produces asymptotic series expansions and a hierarchy of lower-order problems, rather than a single global matrix system~\cite{bayat2012recent}.

The computational cost of solving large nonlinear systems on classical hardware remains a recognized bottleneck, motivating the search for alternative computing paradigms. Quantum algorithms have been developed for nonlinear DEs in conjunction with essentially every classical nonlinearity-treatment strategy discussed above: Euler discretization~\cite{leyton2008quantum}, Dyson series~\cite{berry2024quantum}, Koopman lifting~\cite{joseph2020koopman}, FEM~\cite{montanaro2016quantum}, and Carleman linearization~\cite{lloyd2020quantum}, with the quantum linear system algorithm (QLSA)~\cite{berry2014high, childs2021high, jakhodia2022numerical, krovi2023improved, engel2021linear, akiba2023carleman} typically serving as the underlying linear-solver back-end. These algorithms offer attractive asymptotic complexity but assume fault-tolerant quantum hardware, which is not expected in the near term.

Variational quantum algorithms (VQAs)~\cite{cerezo2021variational} address this near-term gap by employing shallow parameterized circuits whose parameters are optimized by a classical routine to minimize a problem-specific cost function. This hybrid quantum-classical structure provides intrinsic noise resilience and makes VQAs a leading algorithmic family for noisy intermediate-scale quantum (NISQ) devices\cite{liu2024quantum, liu2024VQD}. Within this family, the Variational Quantum Linear Solver (VQLS)~\cite{bravo2019variational} provides a NISQ-compatible back-end for solving linear systems and has seen substantial recent development in performance optimization~\cite{turati2024empirical, kyriienko2021solving, patil2022variational, rao2024performance, hosaka2023preconditioning}, in differentiable circuit formulations for DEs~\cite{lubasch2020variational, tosti2022review, sarma2024quantum}, and in applications to advection--diffusion~\cite{demirdjian2022variational}, Poisson~\cite{cappanera2021variational}, and finite-element problems~\cite{trahan2023variational, arora2025implementation}. Related variational linear-system formulations have broadened the algorithmic landscape further~\cite{xu2021variational, liu2024quantum, liu2024VQD}.

The block structure produced by Carleman linearization is a natural target for both QLSA and VQLS, owing to its sparsity, hierarchical coupling, and amenability to structure-informed ansatz design. To date, however, the integration of Carleman linearization with quantum linear solvers has been confined to quadratic polynomial ODEs, both for the fault-tolerant QLSA pairing~\cite{liu2021efficient, lewis2024limitations, kalmar2025carleman, sanavio2024three, sanavio2024carleman, gonzalez2025quantum} and for the more recent NISQ-compatible VQLS pairing~\cite{surana2024variational}. The extension of this framework to cubic polynomial ODEs---exemplified by the Duffing equation and ubiquitous in nonlinear vibration analysis and structural dynamics---remains, to the best of our knowledge, an open problem.

This paper closes that gap. We formulate the block-banded structure of the cubic Carleman generator and develop a VQLS solver tailored to it, with an ansatz that exploits hierarchical-level coupling. To control the cost of variational measurement, we introduce a symmetry-group strategy based on Hermitianization in the cost function evaluation and treat the non-Hermiticity of the lifted operator via regularized or augmented Hermitianization, thereby controlling the conditioning of the resulting linear system. We benchmark the algorithm against a classical Runge-Kutta reference, and we compare global and local cost function formulations on both the IBM Qiskit~\cite{javadi2024quantum} and Xanadu PennyLane~\cite{bergholm2018pennylane} platforms.

\section{Methods}
\subsection{Carleman Linearization}
\label{sec:carleman}
The Duffing equation belongs to the class of polynomial-in-state ODEs of cubic degree, which admits the canonical Kronecker form
\begin{equation}
    du/dt=F_3u^{\otimes3} + F_1u^{\otimes} + F_0(t), \,\  u(0) = u_{in},
    \label{eq:cubic_polynomial_ode}
\end{equation}
where $u(t) = (u_1(t), \ldots, u_n(t))^{\top} \in \mathbb{R}^n$ is the state vector on the time interval $[0, T]$; $u^{\otimes k} \in \mathbb{R}^{n^k}$ denotes the $k$-fold Kronecker power of $u$; the matrices $F_1 \in \mathbb{R}^{n \times n}$ and $F_3 \in \mathbb{R}^{n\times n^3}$ encode the linear and cubic vector-field coefficients; and $F_0(t) \in \mathbb{R}^n$ represents the (time-dependent) inhomogeneous forcing.

Carleman linearization embeds Eq.~\eqref{eq:cubic_polynomial_ode} into an infinite-dimensional linear flow by promoting the tensor-power coordinates $u^{\otimes k}$ ($k \geq 1$) to independent dynamical variables. Truncating the resulting hierarchy at a finite order $N$ results in the finite-dimensional, linear, time-dependent system
\begin{equation}
    d\hat{y}/dt=A(t)\hat{y}+b(t),\,\  \hat{y}(0)=\hat{y}_{in},
    \label{eq:Carleman_ODEs}
\end{equation}
with lifted state $\hat{y}(t) = \bigl( u^{\otimes 1\top}, \ldots, u^{\otimes N\top} \bigr)^{\top}$, lifted initial condition $\hat{y}_{\mathrm{in}} = \bigl( u_{\mathrm{in}}^{\otimes 1\top}, \ldots, u_{\mathrm{in}}^{\otimes N\top} \bigr)^{\top}$, and inhomogeneity $b(t) = (F_0(t)^{\top}, 0, \ldots, 0)^{\top}$. The generator $A(t)$ inherits a block-banded structure from the polynomial degree of the vector field, with three nonzero block diagonals corresponding to the linear, cubic, and forcing terms, respectively:
\begin{equation}
    \frac{d}{dt}
    \begin{pmatrix}
        \hat y_{1} \\ \hat y_{2} \\ \hat y_{3} \\ \hat y_{4} \\ \vdots \\ \hat y_{N}
    \end{pmatrix}
    =
    \begin{pmatrix}
        A_{1}^{1} & 0         & A_{3}^{1} & 0         & \cdots    & 0         \\
        A_{1}^{2} & A_{2}^{2} & 0         & A_{4}^{2} & \cdots    & 0         \\
        0         & A_{2}^{3} & A_{3}^{3} & 0         & \ddots    & 0         \\
        0         & 0         & A_{3}^{4} & A_{4}^{4} & \ddots    & A_{N}^{N-2}\\
        \vdots    & \vdots    & \ddots    & \ddots    & \ddots    & 0         \\
        0         & 0         & \cdots    & 0         & A_{N-1}^{N} & A_{N}^{N}
    \end{pmatrix}
    \begin{pmatrix}
        \hat y_{1} \\ \hat y_{2} \\ \hat y_{3} \\ \hat y_{4} \\ \vdots \\ \hat y_{N}
    \end{pmatrix}
    +
    \begin{pmatrix}
        F_{0}(t) \\ 0 \\ 0 \\ 0 \\ \vdots \\ 0
    \end{pmatrix}
    \label{eq:Carleman_block_banded}
\end{equation}
The sub-blocks $A_j^j \in \mathbb{R}^{n^j \times n^j}$ (linear coupling), $A_{j+2}^j \in \mathbb{R}^{n^j \times n^{j+2}}$ (cubic coupling), and $A_{j-1}^j \in \mathbb{R}^{n^j \times n^{j-1}}$ (forcing-induced coupling) are obtained by repeated application of the Leibniz rule to $u^{\otimes j}$ and are given by
\begin{equation}
\begin{split}
    &A_{j}^j=\sum_{i=1}^{j} I^{\otimes i-1}\otimes F_1 \otimes I^{\otimes j-i}\\
    &A_{j+2}^j=\sum_{i=1}^{j} I^{\otimes i-1}\otimes F_3 \otimes I^{\otimes j-i}\\
    &A_{j-1}^j=\sum_{i=1}^{j} I^{\otimes i-1} \otimes F_0(t)\otimes I^{\otimes j-i},
\end{split}
\end{equation}
with $A_{j+k}^j = 0$ for $k = 1$ and $k \geq 3$, $A_{j-k}^j = 0$ for $k \geq 2$, and $\hat{y}_j = u^{\otimes j} \in \mathbb{R}^{n^j}$.

To prepare Eq.~\eqref{eq:Carleman_ODEs} for a linear-system formulation, we discretize the time interval $[0, T]$ into $m$ uniform sub-intervals of width $h = T/m$ and apply the first-order explicit (forward) Euler scheme,
\begin{equation}
    y^{k+1}=[I+A(kh)h]y^k + hb(kh),
\end{equation}
where $y^k \in \mathbb{R}^{\Delta}$ approximates $\hat{y}(kh)$ for each $k \in \{0, 1, \ldots, m\}$, $\Delta = n + n^2 + \cdots + n^N = \mathcal{O}(n^N)$ denotes the dimension of the truncated lifted state, and the trajectory is extended by enforcing $y^k = y^m$ for $k \in \{m+1, \ldots, m+p+1\}$. The time step $h$ must be chosen small enough to control the Euler discretization error, and the extension length $p$ large enough to ensure stability of the assembled linear system.

Stacking the Euler recursions for all $k = 0, 1, \ldots, m+p$ into a single linear system yields the block-bidiagonal system $LY = B$:
\begin{equation}
    \begin{pmatrix}
    I & & & & & & & \\
    -[I+A(0)h] & I & & & & & & \\
    & \ddots & \ddots & & & & & \\
    & & -[I+A((m-1)h)h] & I & & & & \\
    & & & -I & I & & & \\
    & & & & \ddots & \ddots & & \\
    & & & & & -I & I
\end{pmatrix}
\begin{pmatrix}
    y^0 \\
    y^1 \\
    \vdots \\
    y^m \\
    y^{m+1} \\
    \vdots \\
    y^{m+p}
\end{pmatrix}
=
\begin{pmatrix}
    y_{in} \\
    b(0) \\
    \vdots \\
    b((m-1)h) \\
    0 \\
    \vdots \\
    0
\end{pmatrix},
\label{eq:Carleman_Euler}
\end{equation}
where $L \in \mathbb{R}^{(m+p+1)\Delta \times (m+p+1)\Delta}$ is a block-bidiagonal matrix whose main block-diagonal is the identity and whose block sub-diagonal contains the propagators $-[I + A(kh)h]$ for $k = 0, \ldots, m-1$ and the identity blocks $-I$ for $k = m, \ldots, m+p-1$. The vector $Y$ collects the trajectory snapshots $\{y^0, y^1, \ldots, y^{m+p}\}$, while the right-hand side $B$ encodes the initial condition $y_{\mathrm{in}}$ together with the discrete forcing $\{b(kh)\}_{k=0}^{m-1}$.

\subsection{Variational Quantum Linear Solver}
\label{sec:VQLS}
The VQLS is a hybrid quantum--classical algorithm for the linear system $L\,|Y\rangle = |B\rangle$, where $L \in \mathbb{C}^{2^Q \times 2^Q}$ and $|B\rangle \in \mathbb{C}^{2^Q}$ on a $Q$-qubit register. The algorithm assumes two ingredients: (i) an efficiently implementable state-preparation unitary $U$ such that $|B\rangle = U\,|0\rangle$~\cite{iten2016quantum, mottonen2004transformation}, and (ii) a linear combination of unitaries (LCU) decomposition~\cite{keen2021quantum, ge2019faster} of $L$ into $n_L$ unitaries $\{L_l\}_{l=1}^{n_L}$,
\begin{equation}
    L = \sum_{l=1}^{n_L} c_l\,L_l, \qquad c_l \in \mathbb{C}.
\end{equation}
A parameterized variational circuit $V(\alpha)$, with classical parameter vector $\alpha$, prepares the trial state $|\psi(\alpha)\rangle = V(\alpha)\,|0\rangle$. Applying $L$ to this trial state and normalizing gives the candidate solution state
\begin{equation}
    |\phi(\alpha)\rangle = L|\psi(\alpha)\rangle /
    \sqrt{\langle \psi(\alpha)|L^{\dagger}L|\psi(\alpha)\rangle},
\end{equation}
and the variational task is to find parameters $\alpha_{\mathrm{opt}}$ such that $L\,|\psi(\alpha_{\mathrm{opt}})\rangle$ is proportional to $|B\rangle$.

Proportionality is enforced by minimizing a cost function that vanishes precisely when $|\phi(\alpha)\rangle$ coincides with $|B\rangle$. With the global Hamiltonian
\begin{equation}
    H_G = L^{\dagger}\bigl(\mathbb{I} - |B\rangle\langle B|\bigr)\,L
\end{equation}
where $\mathbb{I}$ is the identity, the normalized global cost function measures the orthogonal complement of $|\phi(\alpha)\rangle$ with respect to
$|B\rangle$,
\begin{equation}
\begin{split}
    C_G &= \langle \psi | H_G | \psi \rangle / \langle \psi | L^{\dagger} L | \psi \rangle \\
    &= 1 - |\langle B | L | \psi \rangle|^2 / \langle \psi | L^{\dagger} L | \psi \rangle\\
    &= 1 - \sum_{l,l'} c_l\,c_{l'}^{*}\,\gamma_{ll'} / \sum_{l,l'} c_l\,c_{l'}^{*}\,\beta_{ll'},
\end{split} 
\label{eq:cg}
\end{equation}
with the elementary overlaps
\begin{equation}
    \gamma_{ll'}=\langle 0|V^{\dagger} L^{\dagger}_{l'} U |0\rangle \langle 0| U^{\dagger} L_{l} V |0\rangle,
    \label{eq:gamma}
\end{equation}
\begin{equation}
    \beta_{ll'}=\langle 0|V^{\dagger} L^{\dagger}_{l'} L_{l} V |0\rangle,
    \label{eq:beta}
\end{equation}
where $c_{l}^{*}$ denotes the complex conjugate of $c_l$. The overlaps $\gamma_{ll'}$ and $\beta_{ll'}$ are each estimated on the quantum processor via Hadamard Test circuits.

For large $Q$ and sufficiently expressive ansatz circuits, the global cost function suffers from the barren-plateau phenomenon, in which the gradient landscape concentrates exponentially around its mean and trainability degrades~\cite{cerezo2021variational}. A local cost function mitigates this pathology while preserving the equivalence $C_G \to 0 \iff C_L \to 0$~\cite{bravo2019variational}. Replacing the global projector $|0\rangle\langle 0|$ implicit in $\gamma_{ll'}$ (Eq.~\eqref{eq:gamma}) with the qubit-wise local projector
\begin{equation}
    P = 0.5\mathbb{I} + 0.5(\sum_{j=1}^{Q} Z_j)/Q,
\end{equation}
where $Z_j$ is the Pauli-$Z$ operator on qubit $j$, yields
\begin{equation}
    C_L = 0.5- 0.5\sum_{j=1}^{Q}
    \sum_{l,l'} c_l\,c_{l'}^{*}\,\mu^{(j)}_{ll'}/Q\sum_{l,l'} c_l\,c_{l'}^{*}\,\mu^{(-1)}_{ll'}
\label{eq:cl}
\end{equation}
with
\begin{equation}
    \mu^{(j)}_{ll'} = \langle 0 | V^{\dagger} L_{l'}^{\dagger}
   U\,Z_j\,U^{\dagger} L_{l} V | 0 \rangle,
\end{equation}
where the $Z_j$ insertion is implemented in practice as a CZ gate between the ancilla qubit of the Hadamard test and the target qubit $j$.

\subsection{Hadamard Test}
The Hadamard test~\cite{aharonov2006polynomial} is an ancilla-based interference protocol that estimates the real or imaginary part of an inner product $\langle\psi|W|\psi\rangle$, where $|\psi\rangle$ is a normalized state on the $Q$-qubit target register and $W$ is a unitary operator (or a composition of unitaries) acting on the same register. Its shallow circuit depth and reliance on a single ancilla qubit make it well-suited to NISQ-era execution, and we use it as the elementary measurement primitive for the overlaps $\beta_{ll'}$, $\gamma_{ll'}$, and $\mu^{(j)}_{ll'}$ entering the global and local VQLS cost functions (Eqs.~\eqref{eq:cg} and \eqref{eq:cl}).

\begin{figure}[htb!]
    \centering
    \includegraphics[width=\linewidth]{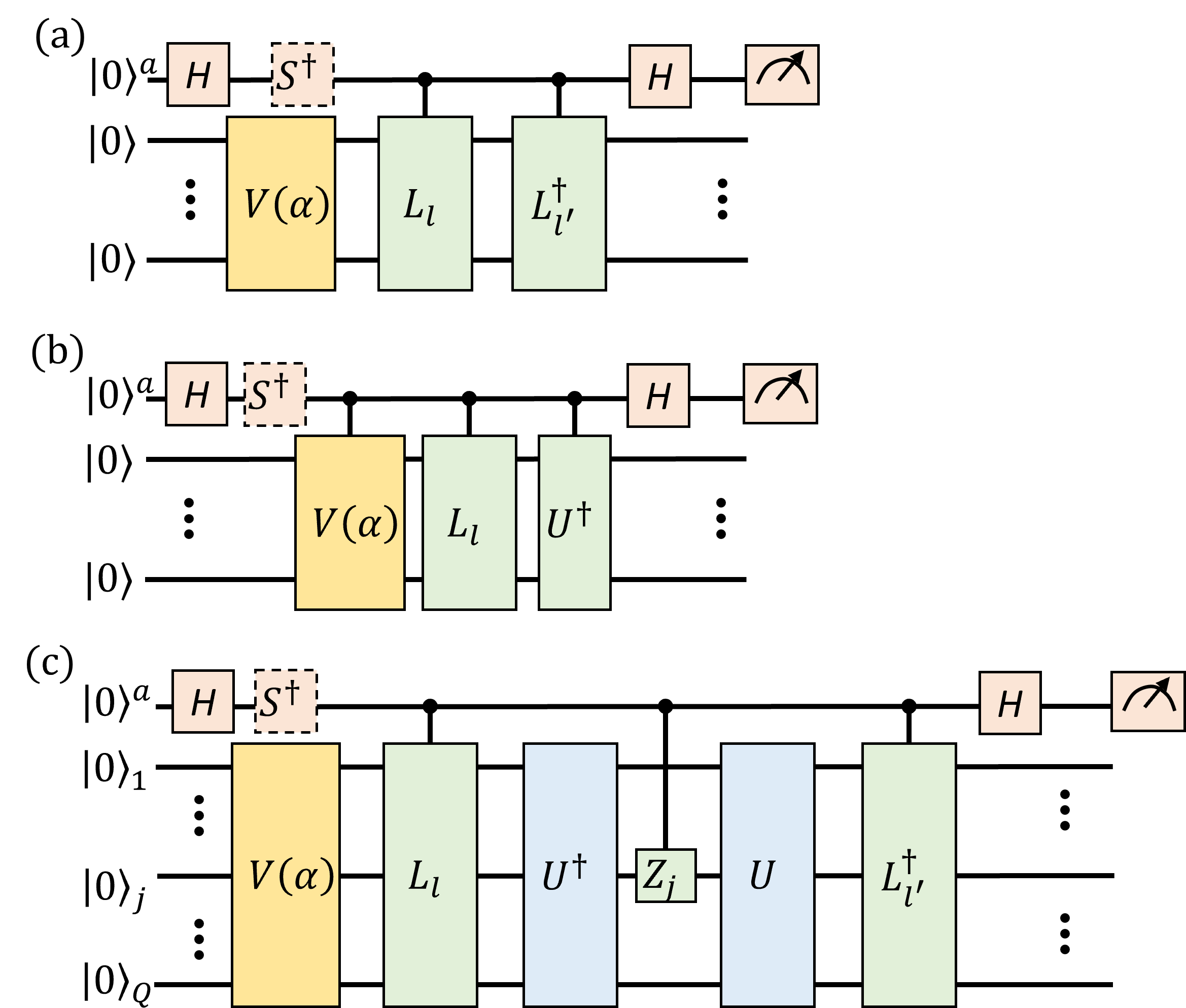}
    \caption{
    Hadamard Test circuits for the three overlap families required by VQLS: (a) $\beta_{ll'}$,     Eq.~\eqref{eq:beta}; (b) $\gamma_{ll'}$,     Eq.~\eqref{eq:gamma}; and (c) $\mu^{(j)}_{ll'}$,     Eq.~\eqref{eq:cl}. The ancilla is denoted $|0\rangle^{a}$; the     target register is $|0\rangle_{1}, \ldots, |0\rangle_{Q}$. Each controlled unitary acts on the target register only when the ancilla is in the $|1\rangle$ state, entangling the conditional branches. The $S^{\dagger}$ gate is inserted only when the imaginary part of the overlap is required; the $Z_j$ insertion in (c) is implemented as a CZ gate between the ancilla and target qubit $j$. Worked circuit examples are provided in the Supplementary Material~\cite{SI}.
    }
    \label{fig:VQLS_circuit}
\end{figure}

Illustrated in Fig.~\ref{fig:VQLS_circuit}, the circuit prepares the ancilla in $|0\rangle$ and the target register in $|\psi\rangle$, applies a Hadamard gate to the ancilla, executes a controlled-$W$ operation conditioned on the ancilla state, applies a second Hadamard to the ancilla, and measures the ancilla in the computational basis. The outcome probabilities $P(0)$ and $P(1)$ have the real part of the target overlap,
\begin{equation}
    \mathrm{Re}\,\langle\psi|W|\psi\rangle = P(0) - P(1).
\label{eq:HT_unified}
\end{equation}
The imaginary part is obtained by inserting an $S^{\dagger}$ phase gate (a $-\pi/2$ rotation about the $z$-axis) on the ancilla between the first Hadamard and the controlled-$W$ operation; the same probability subtraction $P(0) - P(1)$ then returns $\mathrm{Im}\,\langle\psi|W|\psi\rangle$.

In the VQLS implementation, three variants of this circuit are required, one for each overlap family (Fig.~\ref{fig:VQLS_circuit}). For $\beta_{ll'}$, the controlled unitaries are the ansatz $V(\alpha)$ and the LCU operators $L_l$ and $L_{l'}$; for $\gamma_{ll'}$, the controlled state-preparation unitary $U^{\dagger}$ is additionally inserted; and for $\mu^{(j)}_{ll'}$, a controlled-$Z$ on target qubit $j$ is further appended to realize the $Z_j$ projector in the local cost function. The Hadamard Test estimates are combined with the LCU weights $\{c_l\}$ on the classical processor to evaluate Eqs.~\eqref{eq:cg} and \eqref{eq:cl}.

\subsection{Operator Reformulation}
The block-banded operator $L$ assembled in Eq.~\eqref{eq:Carleman_Euler} is generally non-Hermitian and may be ill-conditioned, neither of which is directly compatible with the LCU-based VQLS framework of Section~\ref{sec:VQLS}. Existing strategies for reformulating such operators into VQLS-amenable form include block encoding of non-unitary matrices into probabilistic unitary operations~\cite{gilyen2019quantum}, incomplete-LU (ILU) preconditioning to reduce ansatz depth and improve noise resilience~\cite{hosaka2023preconditioning}, tensorized Pauli decomposition for accelerated multi-qubit operator construction~\cite{hantzko2024tensorized, koska2024tree}, and computational-basis selection for efficient variational implementation~\cite{gnanasekaran2024efficient, de2025variational, znojil2025complex}. These strategies have been benchmarked primarily on the Ising model and random-Pauli operators; to the best of our knowledge, they have not been adapted to the block-banded Carleman-Euler operator structure derived in Section~\ref{sec:carleman}.

We combine three operator reformulation steps tailored to this structure: (i) Hermitianization of $L$, treated in Section~\ref{sec:hermitianization}; (ii) LCU decomposition of the Hermitianized operator into Pauli strings; and (iii) a structure-informed basis selection that exploits the block-banded sparsity. Together, these steps map the assembled Carleman-Euler system onto an efficient VQLS implementation.

\subsubsection{Hermitianization Process}
\label{sec:hermitianization}
We consider two standard Hermitianization schemes for the block-banded $L$, with distinct trade-offs in conditioning and qubit count.
\paragraph{(i) Regularized normal equations} A small diagonal regularization preserves the original Hilbert-space dimension at the cost of approximating the solution:
\begin{equation}
    \bigl(L^{\dagger} L + \epsilon\,\mathbb{I}\bigr)\,Y =
    L^{\dagger}\,B,
    \label{eq:hermitization_normal}
\end{equation}
where $\epsilon$ is a regularization parameter. The condition number of the resulting symmetric positive-definite operator satisfies $\kappa\bigl(L^{\dagger} L + \epsilon\,\mathbb{I}\bigr) = (\sigma_{\max}(L)^2 + \epsilon) / (\sigma_{\min}(L)^2 + \epsilon)$, which scales as $\kappa(L)^2$ for $\epsilon \ll \sigma_{\min}(L)^2$ and saturates at $\sigma_{\max}(L)^2/\epsilon$ as $\sigma_{\min}(L) \to 0$. The scheme therefore controls conditioning at the price of squaring $\kappa(L)$ in the well-conditioned regime.

\paragraph{(ii) Augmented-system dilation} A Hermitian embedding of $L$ into a doubled Hilbert space preserves the exact solution at the cost of one additional qubit,
\begin{equation}
    \begin{pmatrix} 0 & L \\ L^{\dagger} & 0 \end{pmatrix}
    \begin{pmatrix} X \\ Y \end{pmatrix} =
    \begin{pmatrix} B \\ 0 \end{pmatrix},
    \label{eq:hermitization_dilation}
\end{equation}
where $X$ is an ancillary variable that decouples from the target solution $Y$. The dilated operator is Hermitian with singular spectrum $\pm\sigma_k(L)$, so its condition number equals $\kappa(L)$ rather than $\kappa(L)^2$, at the cost of one additional qubit (doubling of the Hilbert-space dimension from $2^Q$ to $2^{Q+1}$).

In both formulations, the resulting Hermitian operator admits a Pauli LCU decomposition with real coefficients, so the elementary overlap  $\mu^{(j)}_{ll'}$ entering the cost function satisfies the Hermitian symmetry $\mu^{(j)}_{l'l} = \overline{\mu^{(j)}_{ll'}}$. Because the LCU coefficients $\{c_l\}$ are real, the imaginary parts of $\mu^{(j)}_{ll'}$ and $\mu^{(j)}_{l'l}$ contribute with opposite signs to the cost function and cancel; only the real, symmetric combinations $\mathrm{Re}\,\mu^{(j)}_{ll'} = \mathrm{Re}\,\mu^{(j)}_{l'l}$ survive. We exploit this symmetry through a symmetry-grouped measurement strategy that reduces the number of distinct Hadamard Test evaluations per cost function call by approximately a factor of two.

\section{Results and Discussion}
\subsection{Carleman-linearized Duffing System}
We first verify the Carleman-Euler assembly of Section~\ref{sec:carleman} on the Duffing equation, Eq.~\eqref{eq:duffing}, by recasting the second-order ODE in first-order state form $u = (z,\,\dot z)^{\top} \in \mathbb{R}^2$, applying Carleman linearization at truncation order $N = 3$, and benchmarking the resulting trajectory against a high-resolution fourth-order Runge-Kutta (RK4) reference. At $N = 3$, the lifted state
spans the $\binom{n+N}{N} - 1 = 9$ non-constant monomials of total
degree $\leq 3$ in the $n = 2$ state variables,
\begin{equation}
    \hat y = \bigl(\,z,\,\dot z,\,z^{2},\,z\dot z,\,\dot z^{2},\,z^{3},\,
    z^{2}\dot z,\,z\dot z^{2},\,\dot z^{3}\,\bigr)^{\top},
    \label{eq:duffing_lifted_state}
\end{equation}
and the forcing enters through the inhomogeneity $b(t) = (0,\,f(t),\,0,\,\ldots,\,0)^{\top}$ with $f(t) = \gamma \cos(\omega t)$, where $\gamma$ is the driving amplitude and $\omega$ the excitation frequency. Assembling the block-banded generator $A(t)$ from Eq.~\eqref{eq:Carleman_block_banded} and applying the forward-Euler discretization of Section~\ref{sec:carleman} produces the block-bidiagonal linear system $L\,Y = B$ of Eq.~\eqref{eq:Carleman_Euler}; the explicit block structure for this case is reported in the Supplementary Material~\cite{SI}.
\begin{figure}[htb!]
    \centering
    \includegraphics[width=\linewidth]{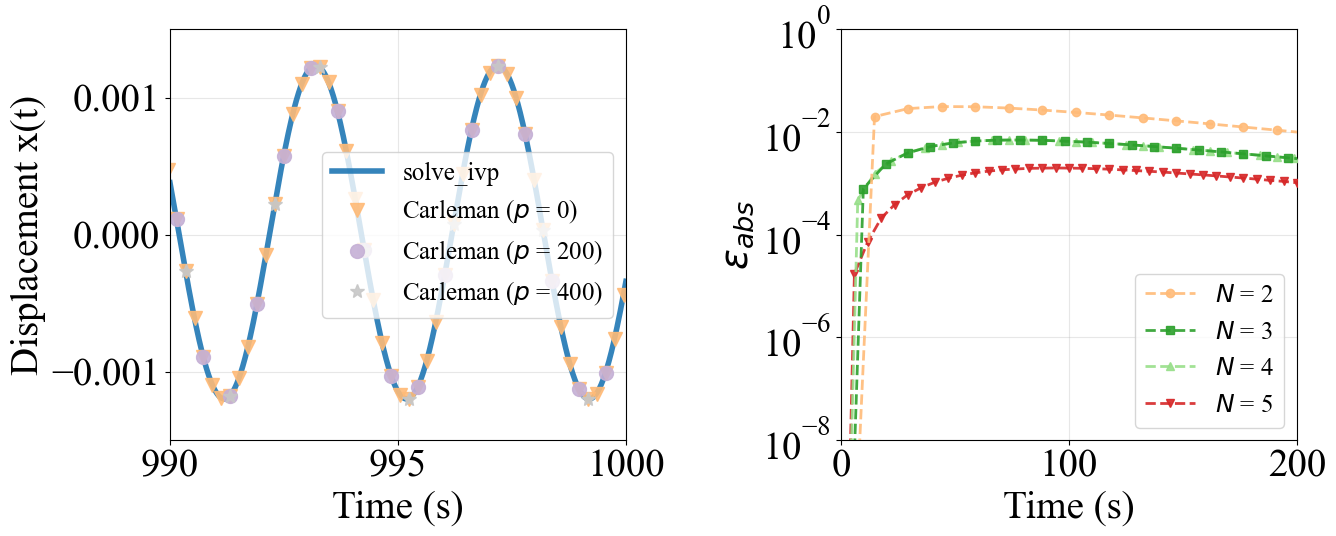}
    \caption{
    Verification of the Carleman linearization on the Duffing equation. (Left) Displacement $z(t)$ from the Carleman method with truncation order $N = 3$, $m = 4 \times 10^{5}$ Euler steps, and stationary-extension length $p \in \{0, 200, 400\}$, compared against the fourth-order Runge-Kutta (RK4) reference. (Right) Pointwise absolute error $|z_{\mathrm{Carleman}}(t) - z_{\mathrm{RK4}}(t)|$ for truncation orders $N \in \{2, 3, 4, 5\}$ with $p = 0$. Hardening-spring parameters: damping $\delta = 5.0\,\mathrm{s^{-1}}$, linear stiffness $\alpha = 0.05\,\mathrm{s^{-2}}$, cubic stiffness $\beta = 0.1\,\mathrm{m^{-2}\,s^{-2}}$, driving amplitude $\gamma = 0.01\,\mathrm{m\,s^{-2}}$, excitation frequency $\omega = 0.5\,\mathrm{rad\,s^{-1}}$, initial displacement $z(0) = 0.5\,\mathrm{m}$, and initial velocity $\dot z(0) = -0.2\,\mathrm{m\,s^{-1}}$. The corresponding convergence ratio~\cite{liu2021efficient} is $R \approx 4.75$ (Eq.~\eqref{eq:R_value}).
    }
    \label{fig:Carleman_periodic}
\end{figure}

Figure~\ref{fig:Carleman_periodic} reports the verification study for this representative hardening-spring regime. The left panel shows that the Carleman-linearized displacement tracks the RK4 reference across the simulation window for all three stationary-extension lengths $p \in \{0, 200, 400\}$, confirming that the trajectory in the active interval $[0, T]$ is insensitive to $p$ (as expected from the construction of Eq.~\eqref{eq:Carleman_Euler}, in which the extended block-rows $k > m$ enforce $y^k = y^m$ and do not propagate dynamics). The right panel reports the pointwise absolute error between the Carleman and RK4 solutions for truncation orders $N \in \{2, 3, 4, 5\}$; the error decreases non-monotonically with $N$.

The convergence ratio~\cite{liu2021efficient},
\begin{equation}
R = (\|z(0)\|^{2}\,\|F_{3}\|
  + \|F_{0}\|_{\infty}/\|z(0)\|)/
 (|\mathrm{Re}\,\lambda_{n}(F_{1})|)
\approx 4.75,
\label{eq:R_value}
\end{equation}
where $|F_{0}\|_{\infty}$ is the peak amplitude of the sinusoidal forcing, $\lambda_{n}(F_{1})$ denotes the eigenvalue of $F_{1}$ with the largest real part, places these parameters outside the dissipative regime $R<1$ originally analyzed for quadratic Carleman linearization in Ref.~\cite{liu2021efficient}. Related work for quadratic systems beyond the dissipative condition is reported in Ref.~\cite{wu2024quantum}. The non-monotone $N$-convergence observed in Fig.~\ref{fig:Carleman_periodic}(right) is therefore presented as an empirical finding for the cubic Duffing case.

\subsection{Numerical Setup for VQLS Benchmarks}
This subsection establishes the numerical setup --- software platforms, cost functions, optimization protocol, ansätze, evaluation metrics, and circuit-depth budget --- used in the VQLS benchmarks reported in the remaining results subsections. The benchmarks target block-banded test systems whose structure mirrors the cubic Carleman-Euler operator of Section~\ref{sec:carleman}, at scales where both quantum and classical verification remain feasible.

\subsubsection{Numerical implementation} 
Two pipelines are implemented in parallel. The IBM Qiskit pipeline uses statevector simulation of the global cost function $C_G$ [Eq.~\eqref{eq:cg}] with the gradient-free COBYLA optimizer~\cite{powell1994direct}, applied to a Pauli-LCU decomposition of the regularized normal equation operator. The Xanadu PennyLane pipeline uses the \texttt{lightning.qubit} statevector backend with the local cost function $C_L$ and an automatic-differentiation-based gradient-descent optimizer~\cite{baydin2018automatic}. Within the PennyLane pipeline, we compare two Hermitianization variants:
\begin{description}
    \item Method A: Pauli-LCU decomposition of the regularized normal equation operator.
    \item Method B: Pauli-LCU decomposition of the augmented-system dilation, combined with the structure-informed basis selection (See in Supplementary Material~\cite{SI}. Method B requires post-selection of the optimized state onto the lower (target) block of the dilation to recover the solution.
\end{description}

\subsubsection{Optimization protocol} All optimizers are run for up to $10^{3}$ iterations with a cost function convergence tolerance of $10^{-8}$. The variational parameters $\alpha$ are initialized from a fixed random seed to ensure reproducibility across runs.

\subsubsection{Ansätze} Two hardware-efficient circuit families are evaluated: a layered Hardware-Efficient Ansatz (HEA) with linear $\mathrm{CNOT}$ entanglement, and a fully ring-entangled ansatz (RING) with cyclic $\mathrm{CNOT}$ entanglement. Both are limited to circuit depth $\leq 5$ layers; explicit circuit diagrams are provided in the Supplementary Material~\cite{SI}.

\subsubsection{Evaluation Metrics}
For each test system, the following quantities are reported, averaged over approximately ten independent runs:
\begin{itemize}
    \item \emph{Input state} $|B\rangle$ --- prepared either with uniformly distributed amplitudes ($b_{\mathrm{uni}}$) or with amplitudes drawn from a fixed random seed; both normalized.
    \item \emph{Condition number} $\kappa(L_H)$ --- the spectral condition number of the Hermitianized operator $L_H$ (regularized normal equation or dilated operator, as appropriate).
    \item \emph{Scaling ratio} $\lambda^{*} = \langle B \,|\, L_{H} \,|\, \psi_{\mathrm{opt}} \rangle$ --- the optimal proportionality between $L_{H}\,|\psi_{\mathrm{opt}}\rangle$ and $|B\rangle$.
    \item \emph{Relative residual} $R_{r} = \bigl\| L_H \,|\,\psi_{\mathrm{opt}} \rangle - \lambda^{*} \,|\, B \rangle \bigr\|/ \| B \|$ --- the post-scaling absolute error.
    \item \emph{Direction fidelity}  $\mathcal{F}_{\mathrm{dir}} = |\langle \hat B \,|\,\widehat{L_H \psi_{\mathrm{opt}}} \rangle|^{2}$ --- the angular alignment between the normalized $L_H\,|\psi_{\mathrm{opt}}\rangle$ and the normalized $|B\rangle$ (hats denote normalization).
    \item \emph{Solution fidelity}
    $\mathcal{F}_{\mathrm{sol}} = |\langle Y_c \,|\, \psi_{\mathrm{opt}}
    \rangle|^{2}$ --- the overlap of the optimal trial state with the
    amplitude-encoded classical reference $|Y_c\rangle$, obtained by
    classical inversion of $L_H Y = B$.
    \item \emph{Bhattacharyya coefficient}
    $\mathrm{BC}(P_c, P_q) = \sum_{i} \sqrt{p^{(c)}_{i}\,p^{(q)}_{i}}$
    --- similarity between the computational-basis probability
    distributions of the classical ($P_c$) and quantum ($P_q$)
    solutions.
    \item \emph{Final cost} $C_{\mathrm{final}}$ --- the cost-function
    value at the last optimizer iteration, used as a convergence
    indicator.
\end{itemize}

\subsubsection{Circuit-depth scaling} 
The cost function circuits scale as
\begin{equation}
    \mathcal{O}\!\left( 4^{N}\,(m+p)^{2}\,
    \bigl(N + \log(m+p)\bigr) \right),
    \label{eq:depth_scaling}
\end{equation}
where the $4^{N}$ factor reflects the exponential growth of the Pauli-LCU term count with truncation order $N$, the $(m+p)^{2}$ factor captures the quadratic growth of Pauli-pair products with the total number of time steps, and the $N + \log(m+p)$ factor is the gate count per controlled-Pauli operator. The augmented-system dilation increases the depth by approximately a factor of four relative to the regularized normal equation (Method A), while the choice between HEA and RING ansätze has a negligible effect on total depth. Two practical consequences follow: (i) tractable circuit synthesis requires keeping the truncation order $N$ small, since the $4^{N}$ growth dominates; and (ii) the quadratic scaling in the time-step count $m$ bounds the dynamical simulation length accessible on near-term hardware. As a concrete benchmark, $m = 30$
and $N = 3$ already require $\sim\!10^{4}$--$10^{5}$ Hadamard Test circuits per cost function evaluation.

\subsection{VQLS Benchmarks on Block-Banded Test Systems}
We benchmark the three pipelines on a sequence of block-banded test systems, sized for both quantum and classical verification within this depth budget, organized by platform and Hermitianization variant.
\begin{figure}[htb!]
    \centering
    \includegraphics[width=\linewidth]{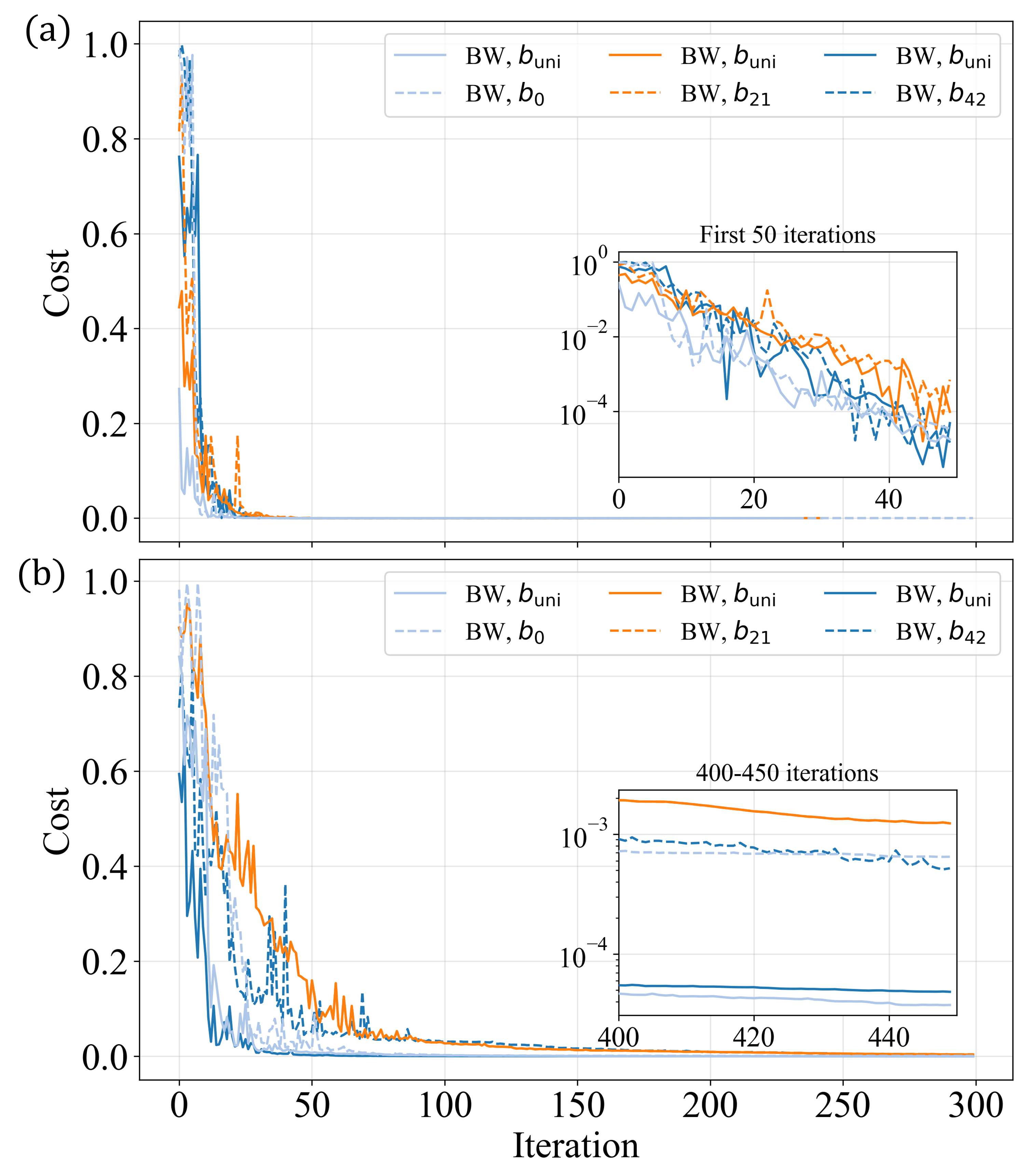}
    \caption{
    Global cost-function history for the Qiskit pipeline (Method A) on (a) a $Q = 2$ and (b) a $Q = 3$ block-banded test system. Insets show the first 50 iterations and iterations 400--450 on a logarithmic scale.
    }
    \label{fig:cost_qiskit}
\end{figure}
\begin{table*}[htb!]
\centering
\caption{VQLS Performance via Qiskit Pipeline}
\adjustbox{width=\textwidth}{
\Large
\renewcommand{\arraystretch}{1.3}
\setlength{\tabcolsep}{8pt}
\begin{tabular}{cccccc|cccccc}
\hline
$Q$ & seed & $B$ & $\kappa(L_H)$ & ansatz & depth & $\lambda^{*}$ & $R_r$ & $\mathcal{F}_{\mathrm{sol}}$ & $\mathcal{F}_{\mathrm{dir}}$ & BC & $C_{G,\mathrm{final}}$ \\
\hline
2 & 0 & $b_{\textrm{seed}}$ & 6.08 & HEA & 2 & -1.9134 & 0.0000 & 1.0000 & 1.0000 & 1.0000 & 3.55e-15 \\
2 & 0 & $b_{\textrm{seed}}$ & 6.08 & HEA & 2 & 2.0628 & 0.0000 & 1.0000 & 1.0000 & 1.0000 & 1.50e-14 \\
2 & 21 & $b_{\textrm{seed}}$ & 2.82 & HEA & 2 & -1.2363 & 0.0000 & 1.0000 & 1.0000 & 1.0000 & 5.22e-15 \\
2 & 21 & $b_{\textrm{seed}}$ & 2.82 & HEA & 2 & 0.8880 & 0.0000 & 1.0000 & 1.0000 & 1.0000 & 3.55e-15 \\
2 & 42 & $b_{\textrm{seed}}$ & 1.97 & HEA & 3 & -1.2910 & 0.0000 & 1.0000 & 1.0000 & 1.0000 & 3.33e-15 \\
2 & 42 & $b_{\textrm{seed}}$ & 1.97 & HEA & 2 & -1.2546 & 0.3047 & 0.8580 & 0.9443 & 0.9263 & 5.57e-02 \\
3 & 0 & $b_{\textrm{seed}}$ & 25.43 & HEA & 2 & -2.0406 & 0.0074 & 0.9975 & 1.0000 & 0.9988 & 1.30e-05 \\
3 & 0 & $b_{\textrm{seed}}$ & 25.43 & HEA & 4 & 1.9939 & 0.0333 & 0.9474 & 0.9997 & 0.9734 & 2.78e-04 \\
3 & 21 & $b_{\textrm{seed}}$ & 8.17 & HEA & 2 & 0.9645 & 0.0076 & 0.9998 & 0.9999 & 0.9999 & 6.16e-05 \\
\hline
3 & 21 & $b_{\textrm{seed}}$ & 8.17 & HEA & 2 & 0.7809 & 0.5661 & 0.4369 & 0.6555 & 0.7702 & 3.45e-01 \\
\hline
3 & 42 & $b_{\textrm{seed}}$ & 5.27 & HEA & 2 & -1.5053 & 0.0052 & 0.9999 & 1.0000 & 0.9999 & 1.19e-05 \\
3 & 42 & $b_{\textrm{seed}}$ & 5.27 & HEA & 4 & 1.2996 & 0.0004 & 1.0000 & 1.0000 & 1.0000 & 7.78e-08 \\
\hline
\hline
\end{tabular}
}
\label{tab:qiskit_table}
\end{table*}

\paragraph{\textbf{Qiskit pipeline (global cost, COBYLA)}} Figure~\ref{fig:cost_qiskit} shows the global cost-function trajectory for representative two- and three-qubit test systems, and Table~\ref{tab:qiskit_table} reports the corresponding solver metrics. In the majority of cases, the optimizer recovers the classical solution to within $\mathcal{F}_{\mathrm{sol}} > 0.99$ and $R_{r} < 10^{-2}$. The highlighted row, however, shows a qualitatively different outcome: the cost has decreased to $C_{G,\mathrm{final}} \approx 0.34$, but the direction fidelity ($\mathcal{F}_{\mathrm{dir}} \approx 0.66$), solution fidelity ($\mathcal{F}_{\mathrm{sol}} \approx 0.44$), and Bhattacharyya coefficient ($\mathrm{BC} \approx 0.77$) remain well below unity. This pattern --- a low cost coupled with poor fidelities --- is consistent with the barren plateau of global cost landscapes for multi-qubit VQLS, in which the cost function becomes exponentially concentrated near its mean, and an optimizer can plausibly converge to a configuration that lies far from the true solution in fidelity space~\cite{cerezo2021variational}.

\paragraph{\textbf{PennyLane pipeline, Method A (local cost, gradient descent; regularized normal equation)}} For both the HEA and RING ansätze, we apply Method A to the same Hermitianized operators $L_{H}$ and Input vectors $B$ used in Table~\ref{tab:qiskit_table}; results are reported in Table~\ref{tab:pl_A_table}. Across all entries, the $\mathcal{F}_{\mathrm{dir}}$, $\mathcal{F}_{\mathrm{sol}}$, and BC are close to unity, and the $R_r$ is close to zero, confirming that $L_{H}\,|\psi_{\mathrm{opt}}\rangle \propto |B\rangle$ at the end of converged optimization. For the challenging instance identified in Table~\ref{tab:qiskit_table}, the local cost function yields substantial improvement under both ansatz families (highlighted rows), recovering $\mathcal{F}_{\mathrm{sol}} \gtrsim 0.999$ where the global cost left it at $C_{L,\textrm{final}}\approx 0.44$. This is consistent with the known mitigation of barren-plateau effects by local cost functions~\cite{bravo2019variational}.

\begin{figure*}[htb!]
    \centering
    \includegraphics[width=.85\linewidth]{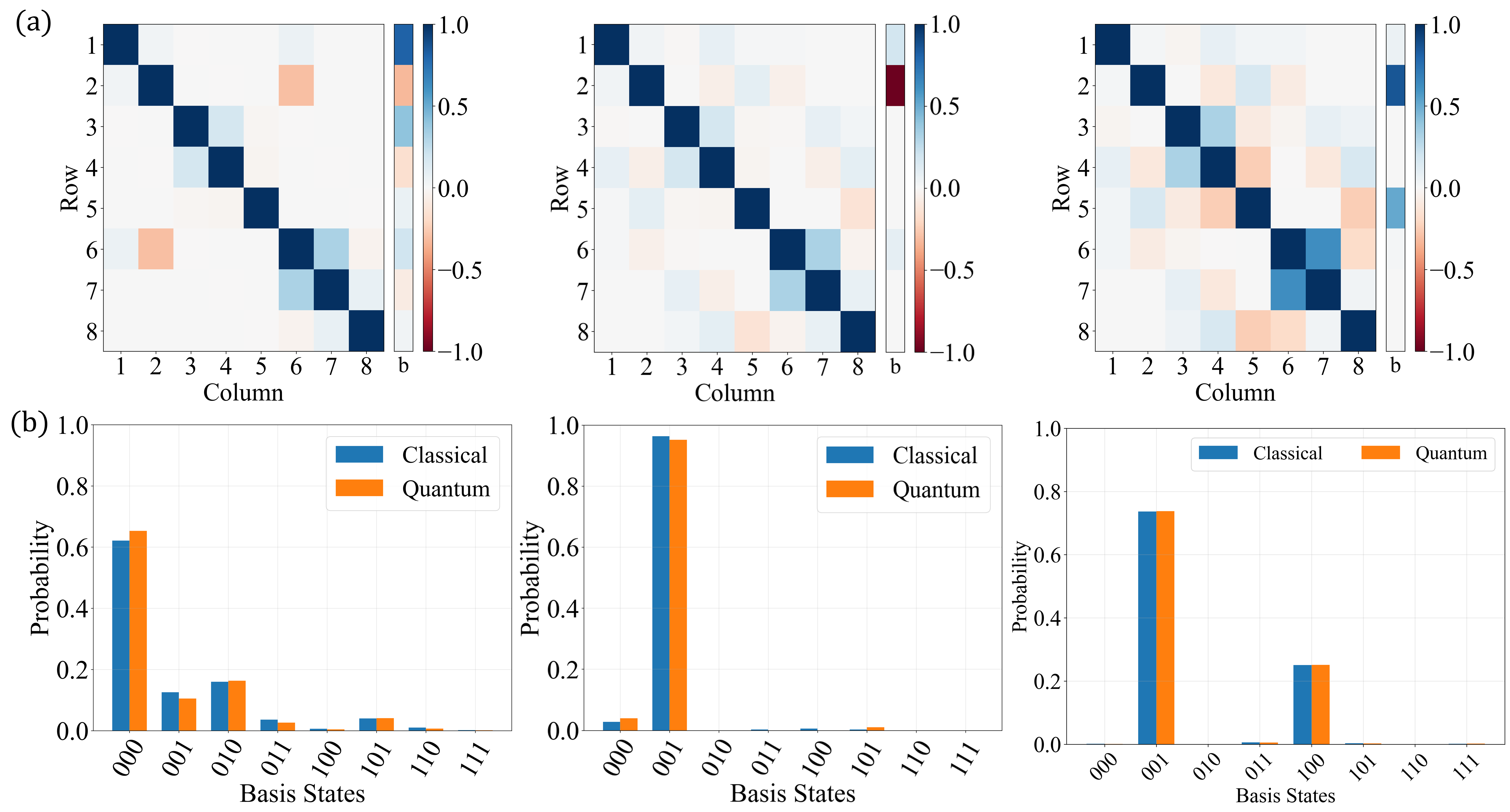}
    \caption{(a) Heatmaps of the regularized system constructed by block-banded matrices and vectors $B$ formulated by the Carleman variables, and (b) probability distributions of quantum and classical solutions.}
    \label{fig:pl_A_vqls}
\end{figure*}

The final block of Table~\ref{tab:pl_A_table} reports
Method A on three Carleman-linearized Duffing systems
(subharmonic, hardening-spring, and superharmonic regimes; parameter sets in the Supplementary Material~\cite{SI}). The corresponding matrices and solution distributions are shown in Fig.~\ref{fig:pl_A_vqls}; fidelities remain high ($\mathcal{F}_{\mathrm{dir}} \geq 0.98$, $\mathrm{BC} \geq 0.93$) across all three regimes.
\begin{table*}[htb!]
\centering
\caption{VQLS Performance via PennyLane Pipeline, Method A}
\adjustbox{width=\textwidth}{
\Large
\renewcommand{\arraystretch}{1.2}
\setlength{\tabcolsep}{8pt}
\begin{tabular}{cccccc|cccccc}
\hline
\hline
$Q$ & seed & $B$ & $\kappa(L_H)$ & ansatz & depth & $\lambda^{*}$ & $R_r$ & $\mathcal{F}_{\mathrm{sol}}$ & $\mathcal{F}_{\mathrm{dir}}$ & BC & $C_{L,\mathrm{final}}$ \\
\hline
2 & 0 & $b_{\textrm{uni}}$ & 6.08 & HEA & 4 & 1.9134 & 0.0000 & 1.0000 & 1.0000 & 1.0000 & 0.00e+00 \\
2 & 0 & $b_{\textrm{seed}}$ & 6.08 & HEA & 4 & -2.0628 & 0.0000 & 1.0000 & 1.0000 & 1.0000 & 7.26e-14 \\
2 & 21 & $b_{\textrm{uni}}$ & 2.82 & HEA & 3 & -1.2363 & 0.0000 & 1.0000 & 1.0000 & 1.0000 & -4.44e-16 \\
2 & 21 & $b_{\textrm{seed}}$ & 2.82 & HEA & 3 & -0.8880 & 0.0000 & 1.0000 & 1.0000 & 1.0000 & 1.11e-16 \\
2 & 42 & $b_{\textrm{uni}}$ & 1.97 & HEA & 4 & -1.2910 & 0.0000 & 1.0000 & 1.0000 & 1.0000 & -2.22e-16 \\
2 & 42 & $b_{\textrm{seed}}$ & 1.97 & HEA & 3 & -1.2171 & 0.0000 & 1.0000 & 1.0000 & 1.0000 & -2.22e-16 \\
3 & 0 & $b_{\textrm{uni}}$ & 25.43 & HEA & 4 & -2.0373 & 0.0074 & 0.9978 & 1.0000 & 0.9989 & 1.31e-05 \\
3 & 0 & $b_{\textrm{seed}}$ & 25.43 & HEA & 3 & 1.9393 & 0.0110 & 0.9950 & 1.0000 & 0.9975 & 3.22e-05 \\
3 & 21 & $b_{\textrm{uni}}$ & 8.17 & HEA & 3 & 0.9589 & 0.0021 & 1.0000 & 1.0000 & 1.0000 & 4.71e-06 \\
\hline
3 & 21 & $b_{\textrm{seed}}$ & 8.17 & HEA & 2 & -1.3013 & 0.0000 & 1.0000 & 1.0000 & 1.0000 & 7.02e-11 \\
\hline
3 & 42 & $b_{\textrm{uni}}$ & 5.27 & HEA & 3 & 1.5024 & 0.0142 & 0.9994 & 0.9999 & 0.9997 & 8.99e-05 \\
3 & 42 & $b_{\textrm{seed}}$ & 5.27 & HEA & 3 & 1.2995 & 0.0000 & 1.0000 & 1.0000 & 1.0000 & 1.83e-10 \\
2 & 0 & $b_{\textrm{uni}}$ & 6.08 & RING & 4 & -1.9134 & 0.0000 & 1.0000 & 1.0000 & 1.0000 & -2.22e-16 \\
2 & 0 & $b_{\textrm{seed}}$ & 6.08 & RING & 3 & 2.0646 & 0.0015 & 1.0000 & 1.0000 & 1.0000 & 5.55e-07 \\
2 & 21 & $b_{\textrm{uni}}$ & 2.82 & RING & 4 & -1.2363 & 0.0000 & 1.0000 & 1.0000 & 1.0000 & -2.22e-16 \\
2 & 21 & $b_{\textrm{seed}}$ & 2.82 & RING & 3 & 0.8880 & 0.0000 & 1.0000 & 1.0000 & 1.0000 & 0.00e+00 \\
2 & 42 & $b_{\textrm{uni}}$ & 1.97 & RING & 4 & -1.2910 & 0.0000 & 1.0000 & 1.0000 & 1.0000 & -2.22e-16 \\
2 & 42 & $b_{\textrm{seed}}$ & 1.97 & RING & 2 & 1.2171 & 0.0000 & 1.0000 & 1.0000 & 1.0000 & 0.00e+00 \\
3 & 0 & $b_{\textrm{uni}}$ & 25.43 & RING & 3 & 2.0317 & 0.0033 & 0.9995 & 1.0000 & 0.9998 & 2.61e-06 \\
3 & 0 & $b_{\textrm{seed}}$ & 25.43 & RING & 3 & 1.9286 & 0.0084 & 0.9972 & 1.0000 & 0.9986 & 1.88e-05 \\
3 & 21 & $b_{\textrm{uni}}$ & 8.17 & RING & 3 & 0.9564 & 0.0114 & 0.9992 & 0.9999 & 0.9996 & 1.42e-04 \\
\hline
3 & 21 & $b_{\textrm{seed}}$ & 8.17 & RING & 3 & 1.2541 & 0.0087 & 0.9998 & 1.0000 & 1.0000 & 5.44e-10 \\
\hline
3 & 42 & $b_{\textrm{uni}}$ & 5.27 & RING & 3 & -1.5057 & 0.0027 & 1.0000 & 1.0000 & 1.0000 & 3.18e-06 \\
3 & 42 & $b_{\textrm{seed}}$ & 5.27 & RING & 4 & -1.2995 & 0.0000 & 1.0000 & 1.0000 & 1.0000 & 1.24e-13 \\
\hline
\hline
$Q$ & $N$ & \multicolumn{1}{c}{Carleman} & \multicolumn{1}{c}{$\kappa(L_H)$} & ansatz & depth & 
$\lambda^{*}$ & $R_r$ & 
$\mathcal{F}_{\mathrm{sol}}$ & 
$\mathcal{F}_{\mathrm{sol}}$ & BC & $C_{L,\mathrm{final}}$ \\
\hline
3 & 3 & subharmonic & 1.26 & HEA & 6 & -1.9979 & 0.0361 & 0.9996 & 0.9997 & 0.9998 & 3.27e-04 \\
3 & 3 & hardening\_spring & 1.85 & HEA & 3 & 1.8642 & 0.2562 & 0.9729 & 0.9815 & 0.9864 & 1.85e-02 \\
3 & 3 & hardening\_spring & 1.29 & HEA & 8 & -1.9568 & 0.0966 & 0.8733 & 0.9976 & 0.9345 & 2.43e-03 \\
3 & 3 & hardening\_spring & 1.29 & HEA & 6 & -1.9569 & 0.0965 & 0.8737 & 0.9976 & 0.9347 & 2.43e-03 \\
3 & 3 & superharmonic & 2.00 & HEA & 8 & -1.9644 & 0.2531 & 0.9793 & 0.9837 & 0.9896 & 1.63e-02 \\
3 & 3 & superharmonic & 1.86 & HEA & 6 & -1.6064 & 0.1594 & 0.9904 & 0.9903 & 0.9952 & 9.75e-03 \\
\hline
\hline
\end{tabular}
}
\label{tab:pl_A_table}
\end{table*}

\begin{figure*}[htbp]
    \centering
    \includegraphics[width=\linewidth]{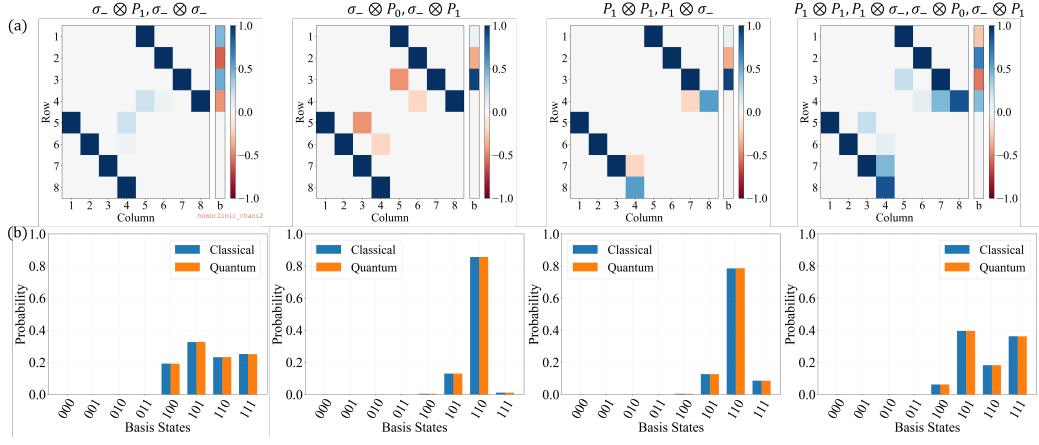}
    \caption{(a) Heatmaps of the augmented system constructed by real-valued block-banded matrices and seed-generated vectors $b_{\textrm{seed}}$, with the structure-informed basis shown on top (See in \textit{Supplementary Material}\cite{SI}, and (b) probability distributions of quantum and classical solutions.}
    \label{fig:pl_B_vqls}
\end{figure*}

\paragraph{\textbf{PennyLane pipeline, Method B (local cost, gradient descent; augmented-system dilation)}} Figure~\ref{fig:pl_B_vqls}(a) depicts the dilated operators, and panel (b) the full probability distributions of the quantum and classical solutions; the latter show high visual similarity even before post-selection. For consistency across rows of Table~\ref{tab:pl_B_table}, we report the unfiltered distributions; in deployment, post-selection onto the lower (target) block of the dilation is required to recover the solution $|Y\rangle$ from the full augmented state.

For the challenging instance from Tables~\ref{tab:qiskit_table} and \ref{tab:pl_A_table}, the highlighted entry of Table~\ref{tab:pl_B_table} shows that Method B achieves $\mathcal{F}_{\mathrm{dir}} = 1.000$ and $\mathrm{BC} = 0.972$, matching or exceeding Method A on these two metrics. The solution fidelity $\mathcal{F}_{\mathrm{sol}} = 0.064$ in the same row is low, not because of a solver failure, but because the augmented state lives in a $2^{Q_{\mathrm{aug}}}$-dimensional space containing both the target block (which carries the solution) and an auxiliary block; the fidelity against the classical reference, which lives in the target block alone, is therefore mechanically suppressed by the unfiltered support on the auxiliary block. Post-selection restores the solution fidelity, as documented in the additional cases in the Supplementary Material~\cite{SI}.
\begin{table*}[htb!]
\centering
\caption{VQLS Performance via PennyLane Pipeline, Method B}
\adjustbox{width=\textwidth}{
\Large
\renewcommand{\arraystretch}{1.3}
\setlength{\tabcolsep}{8pt}
\begin{tabular}{cccccc|cccccc}
\hline
\hline
$Q_{\textrm{aug}}$ & seed & $B$ & $\kappa(L_H)$ & ansatz & depth & $\lambda^{*}$ & $R_r$ & $\mathcal{F}_{\mathrm{sol}}$ & $\mathcal{F}_{\mathrm{dir}}$ & BC & $C_L$[-1] \\
\hline
4 & 21 & $b_{\textrm{seed}}$ & 1.64 & HEA & 2 & 0.9210 & 0.0000 & 0.0638 & 1.0000 & 0.9720 & 2.21e-11 \\
\hline
3 & 0 & $b_{\textrm{seed}}$ & 1.25 & HEA & 3 & -0.9684 & 0.0020 & 0.0042 & 1.0000 & 0.9842 & 4.43e-06 \\
3 & 21 & $b_{\textrm{seed}}$ & 1.94 & HEA & 3 & 0.9568 & 0.0000 & 0.5647 & 1.0000 & 0.9960 & 3.59e-11 \\
3 & 21 & $b_{\textrm{seed}}$ & 1.56 & HEA & 3 & 0.9686 & 0.0000 & 0.5263 & 1.0000 & 0.9813 & 2.42e-11 \\
3 & 42 & $b_{\textrm{seed}}$ & 1.75 & HEA & 4 & 0.9204 & 0.0000 & 0.0935 & 1.0000 & 0.9024 & 2.41e-11 \\
4 & 21 & $b_{\textrm{seed}}$ & 2.60 & HEA & 4 & 0.9566 & 0.0000 & 0.5649 & 1.0000 & 0.9920 & 3.59e-11 \\
4 & 42 & $b_{\textrm{seed}}$ & 2.27 & HEA & 4 & 0.9858 & 0.0000 & 0.5547 & 1.0000 & 0.9961 & 3.48e-11 \\
4 & 42 & $b_{\textrm{seed}}$ & 2.56 & HEA & 4 & 0.9672 & 0.0000 & 0.5366 & 1.0000 & 0.9803 & 1.48e-11 \\
\hline
\end{tabular}
}
\label{tab:pl_B_table}
\end{table*}

Across all three configurations, the structured linear systems obtained by Carleman linearization admit accurate VQLS solutions that are proportional to the classical reference. The combination of regularized Hermitianization with a local cost function (PennyLane Method A) provides the strongest single-pipeline performance on the tested block-banded systems, while the augmented-system dilation (Method B) offers high direction fidelity at the cost of an additional post-selection step and one extra qubit.

\section{Conclusion}
We have developed a hybrid quantum-classical algorithm that couples Carleman linearization with the Variational Quantum Linear Solver (VQLS) for cubic polynomial ordinary differential equations, with the damped, externally forced Duffing oscillator as the canonical benchmark. The contribution is twofold: (i) a verified Carleman-Euler assembly of the cubic Duffing equation that produces a block-banded linear system suitable for variational quantum linear solvers; and (ii) a structure-informed VQLS implementation on this system, comparing two Hermitianization variants --- the regularized normal equation and the augmented-system dilation --- across the IBM Qiskit and Xanadu PennyLane platforms.

On the Carleman side, the absolute error against the fourth-order Runge-Kutta reference decreases monotonically with the truncation order $N \in \{2, 3, 4, 5\}$ and is insensitive to the stationary-extension length $p \in \{0, 200, 400\}$ within the active simulation window, consistent with the construction of Eq.~\eqref{eq:Carleman_Euler}. This non-monotone $N$-convergence is observed even at parameters that place the system outside Liu et al.'s dissipative regime $R < 1$, providing empirical evidence for the cubic case where rigorous error bounds are not yet available. On the VQLS side, the global cost framework (Qiskit pipeline; regularized normal equation; LCU; COBYLA) can drive the cost to near zero while leaving the direction and solution fidelities far from unity --- a signature of the cost concentration of global cost landscapes in multi-qubit VQLS. The local cost framework (PennyLane pipeline; gradient descent), by contrast, consistently yields high direction fidelity, high solution fidelity, and high Bhattacharyya coefficient with low relative residual under both the HEA and RING ansätze. The augmented-system dilation (Method B) further improves direction fidelity and distribution similarity at the cost of one additional qubit and a post-selection step onto the target block of the dilation.

Several extensions of this work are natural. First, we plan to develop the symmetry-based measurement strategy of Section~\ref{sec:hermitianization} into a more general, structure-aware measurement protocol, aiming to further reduce the Hadamard Test count for larger block-banded systems. Second, we intend to construct VQLS circuits via unitary dilations of non-unitary operators (see Supplementary Material~\cite{SI}) to reduce circuit depth. Third, we will apply block encoding to non-Hermitian operators arising in wave-dynamics applications --- including acoustic metamaterials with absorption, topological sound insulators with losses, and active acoustic devices with balanced gain and loss --- in which the linearized operator is intrinsically non-Hermitian. Finally, we plan a noise-resilience analysis of the Carleman--VQLS pipeline using adaptive circuits and mid-circuit measurement, with the goal of extending the framework to dissipative nonlinear partial differential equations at the $40$--$45$ qubit scale accessible on GPU-accelerated quantum simulators.

\section{CRediT author statement}
\textbf{Yunya Liu}: Conceptualization, Methodology, Software, Validation, Formal analysis, Investigation, Data Curation, Writing - Original Draft, Writing - Review \& Editing, Visualization. \textbf{Pai Wang}: Conceptualization, Methodology, Resources, Writing - Review \& Editing, Supervision, Project administration, Funding acquisition.

\section{Declaration of competing interests}
The authors declare that they have no known competing financial interests or personal relationships that could have appeared to influence the work reported in this paper.
\section{Code Availability}
The code that supports the findings of this study is available from the corresponding author upon reasonable request.

\section{Acknowledgment}
YL and PW are supported by both the Research Incentive Seed Grant Program and the start-up research funds of the Department of Mechanical Engineering at the University of Utah. The support and resources from the Center for High-Performance Computing at the University of Utah and the qBraid computing platform are gratefully acknowledged.

\bibliographystyle{elsarticle-num} 
\bibliography{ref}

\end{document}


\begin{frontmatter}
\title{Supplementary Material\\ to\\ Measurement-Efficient Variational Quantum Linear Solver for Carleman-Linearized Nonlinear Dynamics}

\author[gt]{Yunya Liu}
\address[gt]{Woodruff School of Mechanical Engineering,
             Georgia Institute of Technology, Atlanta, Georgia, USA}
\author[ut]{Pai Wang\corref{cor1}}
\ead{pai.wang@utah.edu}
\address[ut]{Department of Mechanical Engineering, University of Utah, Salt Lake City, UT, USA}
\cortext[cor1]{Corresponding author}
\end{frontmatter}

\newpage
\tableofcontents
\clearpage
\section{Carleman-linearized Duffing system}
\begin{figure}[htbp]
    \centering
    \includegraphics[width=\linewidth]{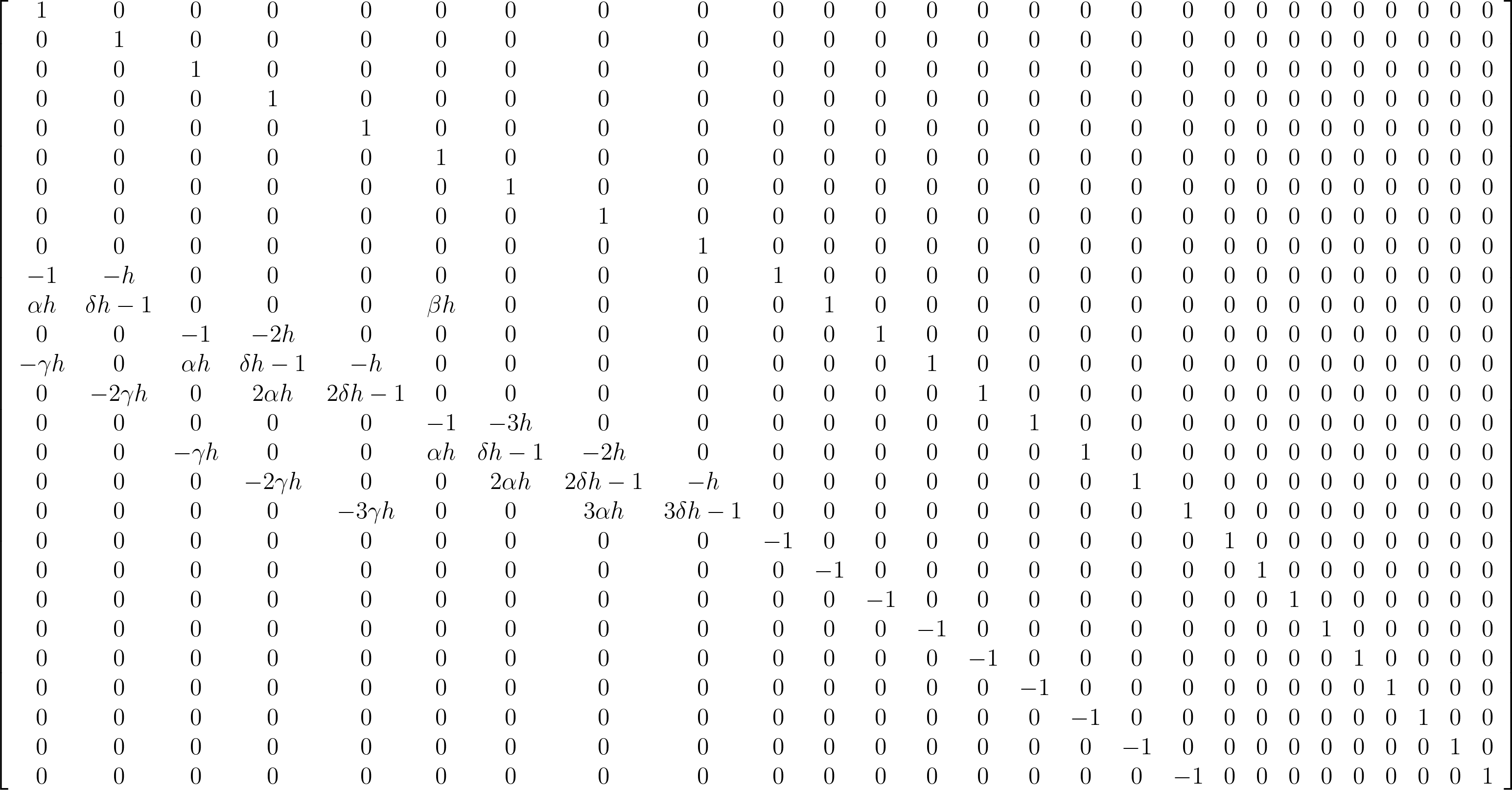}
    \caption{The symbolic form of matrix $L$ where step $m=1$, extended step $p=1$, $h$ is time step, $\delta$ is damping coefficient, $\alpha$ is linear stiffness coefficient, $\beta$ is nonlinear stiffness coefficient, and $\gamma$ is driving amplitude.}
    \label{fig:symbolic_L}
\end{figure}

The Fig.\,\ref{fig:symbolic_L} displays an example of the matrix $L$ with two steps of $m=1$ and $p=1$, which is a sparse block lower-bidiagonal matrix.
\newpage
\section{Hadamard Test Circuits}
Here, we present examples of Hadamard Test circuits using different variational ansätze and controlled unitaries based on the local cost function.
\begin{figure}[htb!]
    \centering
\includegraphics[width=\linewidth]{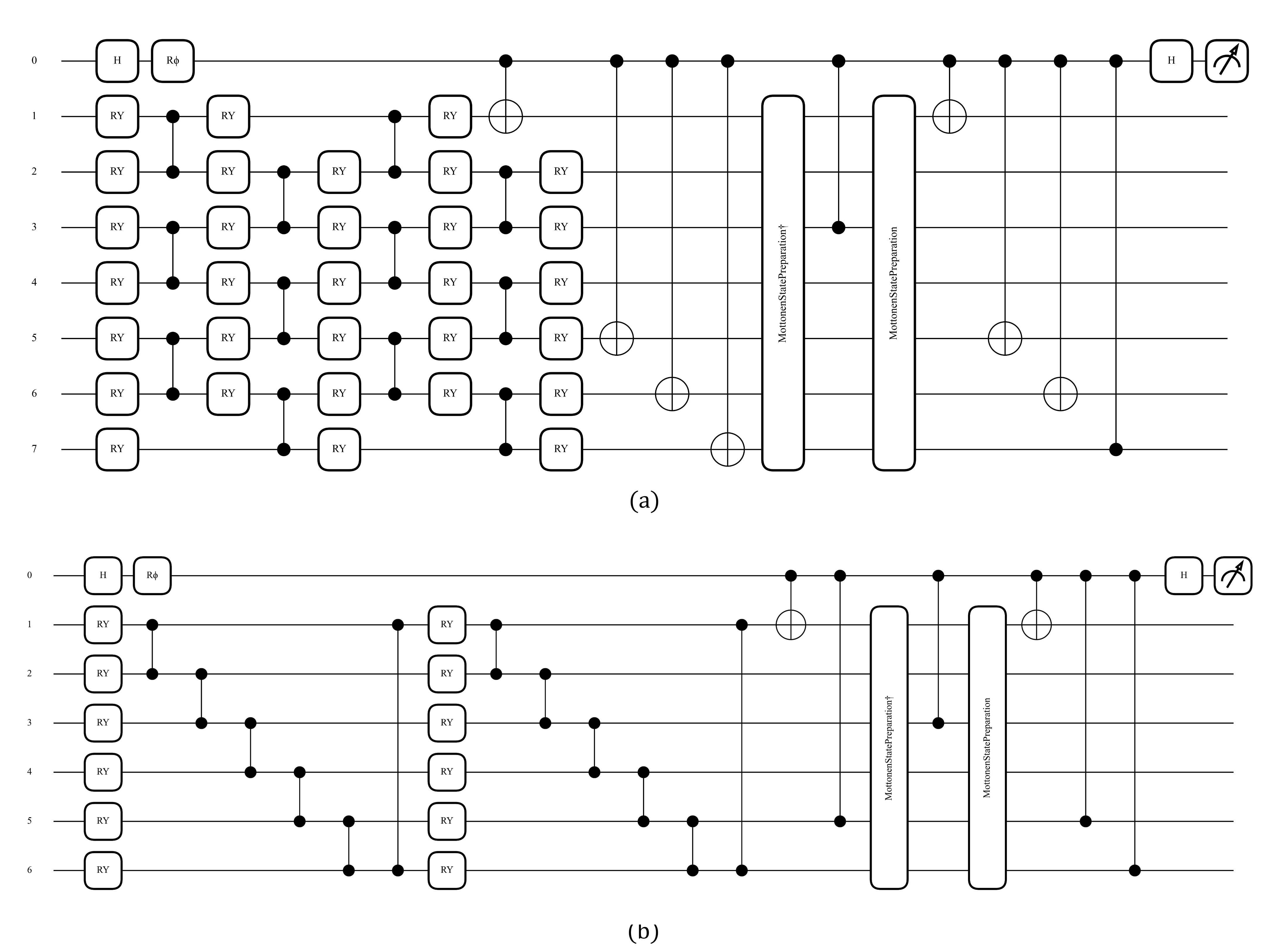}
    \caption{The Hadamard Test circuits for the imaginary part of expectation value measurement. (a) On six target qubits and one ancilla qubit, there are two circuit depths of Layered Hardware-Efficient Ansatz (HEA), controlled Pauli term $L_l=X_0X_4X_5X_6$, $L_{l'}=X_0X_4X_5Z_6$, controlled unitary $Z_2$, and MottonenStatePreparation $U$ for vector $|B\rangle$. (b) On five target qubits and one ancilla qubit, there are two circuit depths of full ring entangled ansatz (RING), controlled Pauli term $L_l=X_0Z_4$, $L_{l'}=X_0Z_4Z_5$, and controlled unitary $Z_2$, and MottonenStatePreparation $U$ for vector $|B\rangle$.
    }
    \label{fig:PL_Vs}
\end{figure}
To explicitly visualize the commuting groups of Hamiltonians (i.e., Pauli operators), we display Fig.\,\ref{fig:PL_commute_Lg} with two distinct examples of $L_g$.
\begin{figure}[htb!]
    \centering
    \includegraphics[width=\linewidth]{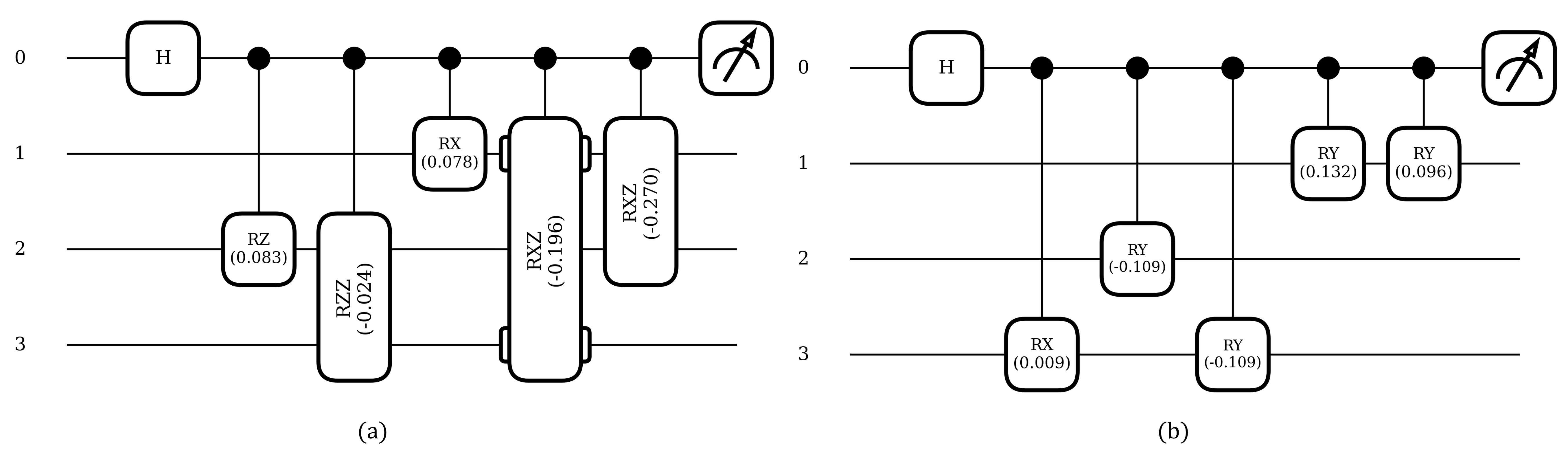}
    \caption{The commuting group strategy for a system with three target qubits and one ancilla qubit. (a) Compared to individual $L_l$ or $L_{lp}$, the grouped Pauli operator $L_g=0.083Z_1-0.024Z_1Z_2+0.078X_0-0.196X_0Z_2-0.270X_0Z_1$ contains five commuting terms. (b) A more explicit visualization shows that four individual Pauli operators are grouped into $L_g=0.009X_2-0.109Y_1Y_2+0.132Y_0+0.096Y_0$. The angles in radians are shown on the circuit as $\alpha = 2Re(c_i)t_{\textrm{evolve}}$ where $t_{\textrm{evolve}}=1.0$. }
    \label{fig:PL_commute_Lg}
\end{figure}\\

\newpage
\section{Appendix for PennyLane Method A}
Here, we present an example of a block-banded matrix applied in the PennyLane Method A approach. We conduct the linear combination of unitaries (LCU) using the Pauli basis $\{I, X, Y, Z\}$.
\begin{figure}[htbp]
    \centering
    \includegraphics[width=\linewidth]{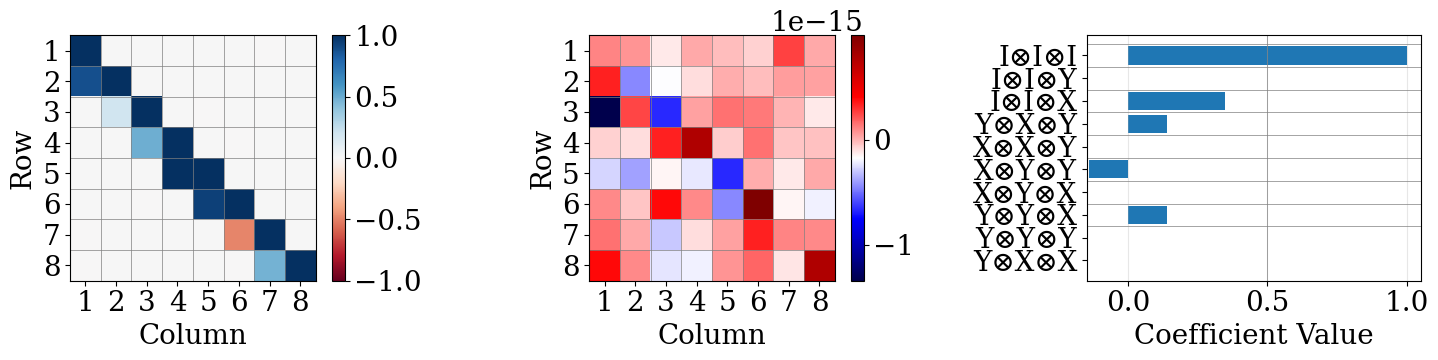}
    \caption{The block-banded matrix used in PennyLane Method A approach: (left) the block-banded target matrix example with banded pattern, (middle) error between reconstructed and target matrices, and (right) top 10 tensor products of Pauli basis corresponding to highest absolute coefficients.}
    \label{fig:pauli_lower_tri}
\end{figure}
\begin{table}[htbp]
\centering
\caption{Duffing Oscillator Parameters}
\renewcommand{\arraystretch}{1.6}
\setlength{\tabcolsep}{8pt}
\begin{tabular}{lccccccc}
\hline
\hline
\textbf{Configuration} & $\delta$ & $\alpha$ & $\beta$ & $\gamma$ & $\omega$ & $x_0$ & $v_0$ \\
\hline
hardening\_spring & 2.0 & 1.0 & 3.0 & 0.01 & $0.5\pi$ & 0.5 & -0.2 \\
\hline
superharmonic & 2.0 & 1.0 & 0.5 & 0.8 & 0.33 & 0.0 & 0.0 \\
\hline
superharmonic & 2.0 & 1.0 & 3.0 & 1.2 & 0.33 & 0.05 & 0.02 \\
\hline
\hline
\end{tabular}
\label{tab:duff_parameters}
\end{table}\\
Table \,\ref{tab:duff_parameters} shows the parameters for Fig.5 in the main text.

\newpage
\section{Appendix for PennyLane Method B}
For $Q$ number of qubits, the Pauli basis consists of all possible tensor products of single-qubit Pauli matrices, forming linearly independent matrices that completely span the $4^Q$-dimensional space of $2^Q \times 2^Q$ complex matrices.\\
The Sigma basis has two options and also generates $4^Q$ linearly independent matrices:
\begin{itemize}
    \item $\{I, Z, \sigma_{+}, \sigma_{-}\}$: It combines the diagonal Pauli operators ($I$ and $Z$) with the ladder operators ($\sigma_{+}$ and $\sigma_{-}$), providing a natural decomposition that separates diagonal elements from off-diagonal transitions.
    \item $\{P_0, P_1, \sigma_{+}, \sigma_{-}\}$: It uses the complete set of projection operators ($P_0=|0\rangle\langle0|$ and $P_1=|1\rangle\langle1|$) along with ladder operators ($\sigma_{+}=|0\rangle\langle1|$ and $\sigma_{-}=|1\rangle\langle0|$), representing all possible single-qubit quantum transitions and projections.
\end{itemize}
\begin{figure}[htbp]
    \centering
    \includegraphics[width=\linewidth]{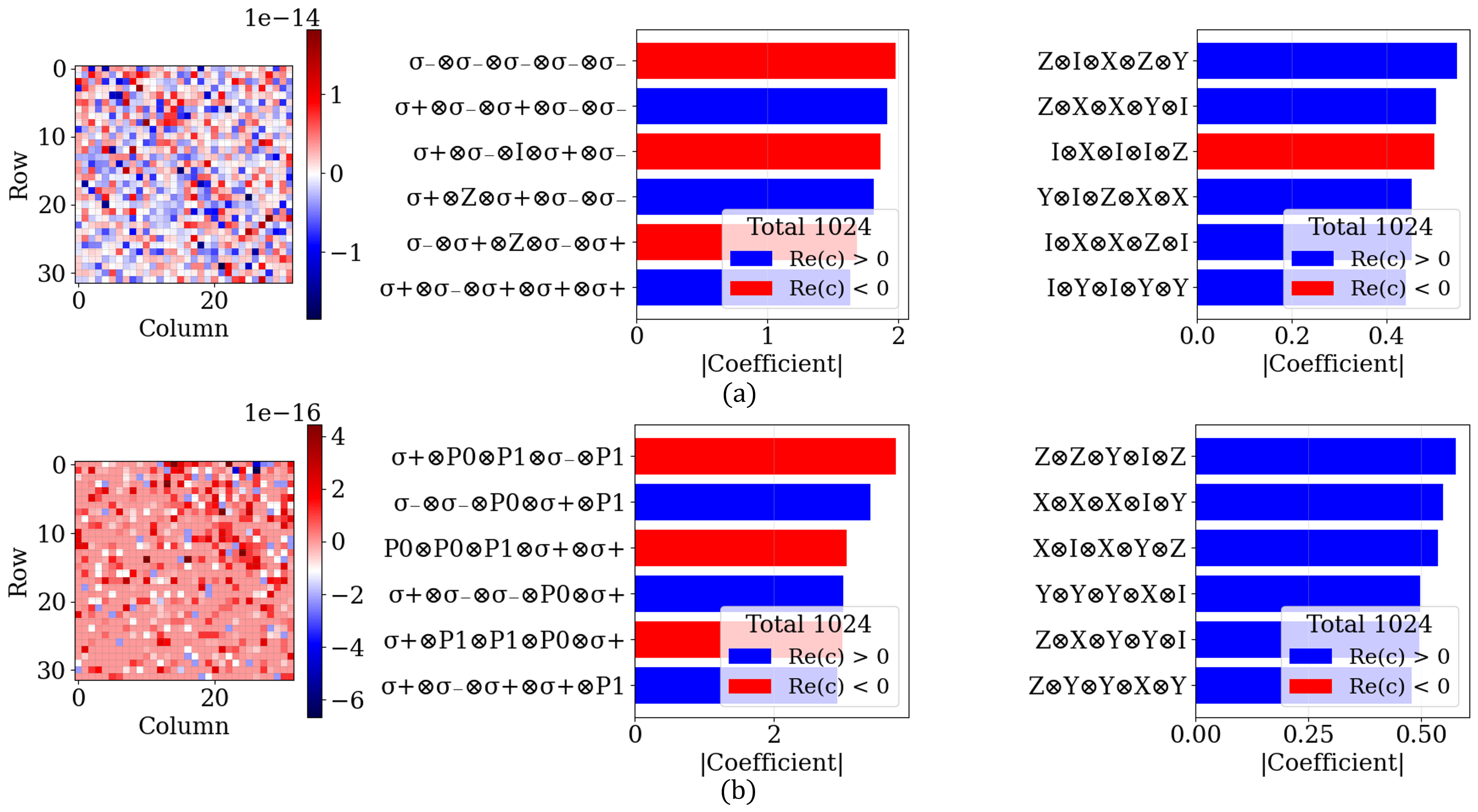}
    \caption{Comparison of different basis for two real-valued block-banded matrix in (a) and (b): (first column) error between a target matrix and the reconstructed matrix using Pauli basis of, (second column) absolute coefficients distribution of Sigma basis $\{I, Z, \sigma_{+}, \sigma_{-}\}$ (top) and $\{P0, P1, \sigma_{+}, \sigma_{-}\}$ (bottom), and (last column) absolute coefficients distribution of Pauli basis.}
    \label{fig:sigma_vs_pauli}
\end{figure}
As shown in Fig.\,\ref{fig:pl_b_Q3} and \ref{fig:pl_b_Q4}(a), the cost history of examples applied in the main text shows convergence beneath $10^{-4}$ for $Q = 3$ and $10^{-2}$ for $Q = 4$. The number of qubits $Q_{\textrm{aug}}$ is from the augmented system, $b_{\textrm{seed}}$ is a vector generated by the seed, and depth is the repeated number of layered Hardware-Efficient Ansatz (HEA).
\begin{figure}[htbp]
    \centering
    \includegraphics[width=.65\linewidth]{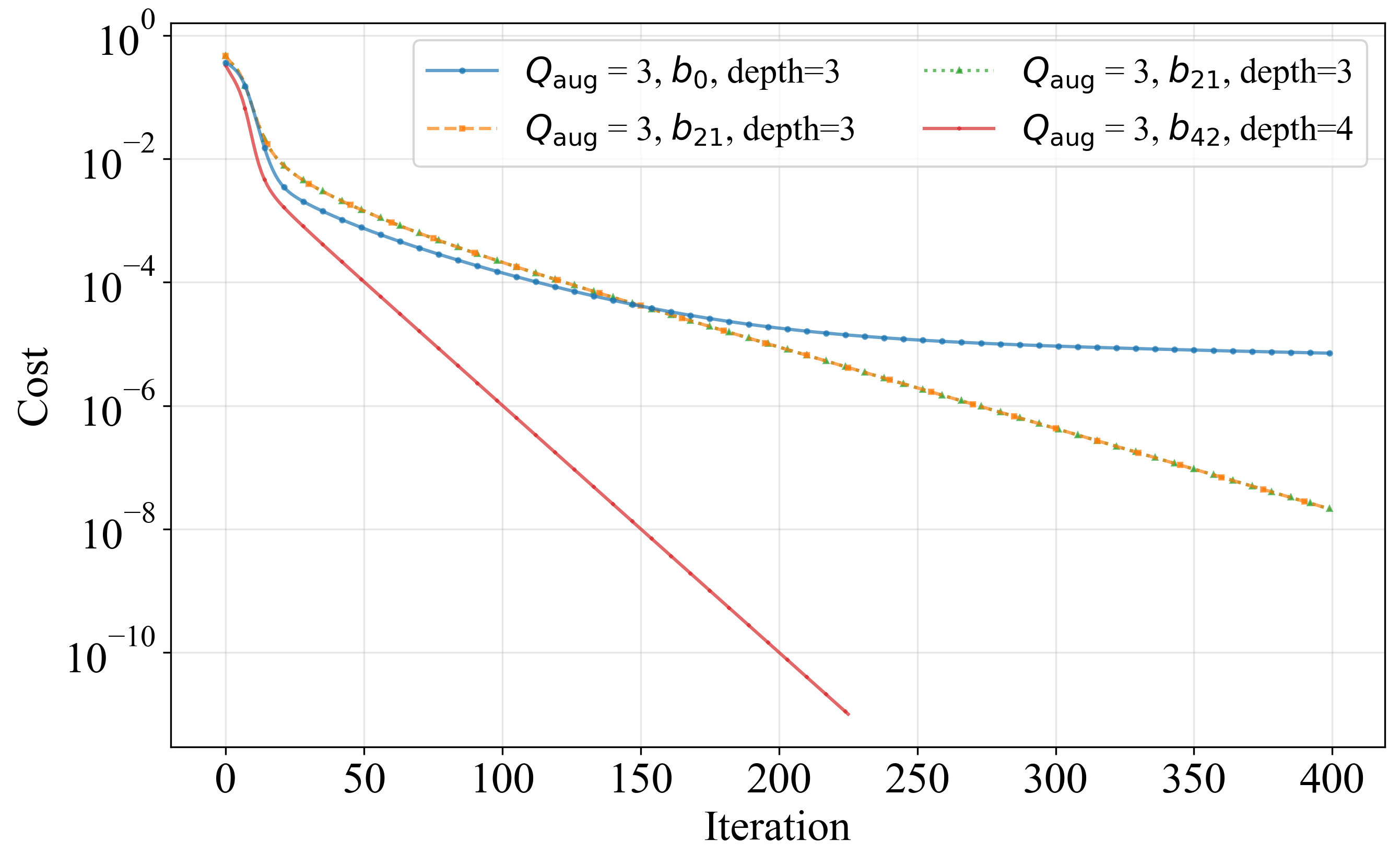}
    \caption{Cost history of examples of real-valued block-banded matrices generated by the Sigma basis. Qubit number $Q = 3$ corresponds to the first four entries in Table III of the main text.}
    \label{fig:pl_b_Q3}
\end{figure}
\begin{figure}[htbp]
    \centering
    \includegraphics[width=.8\linewidth]{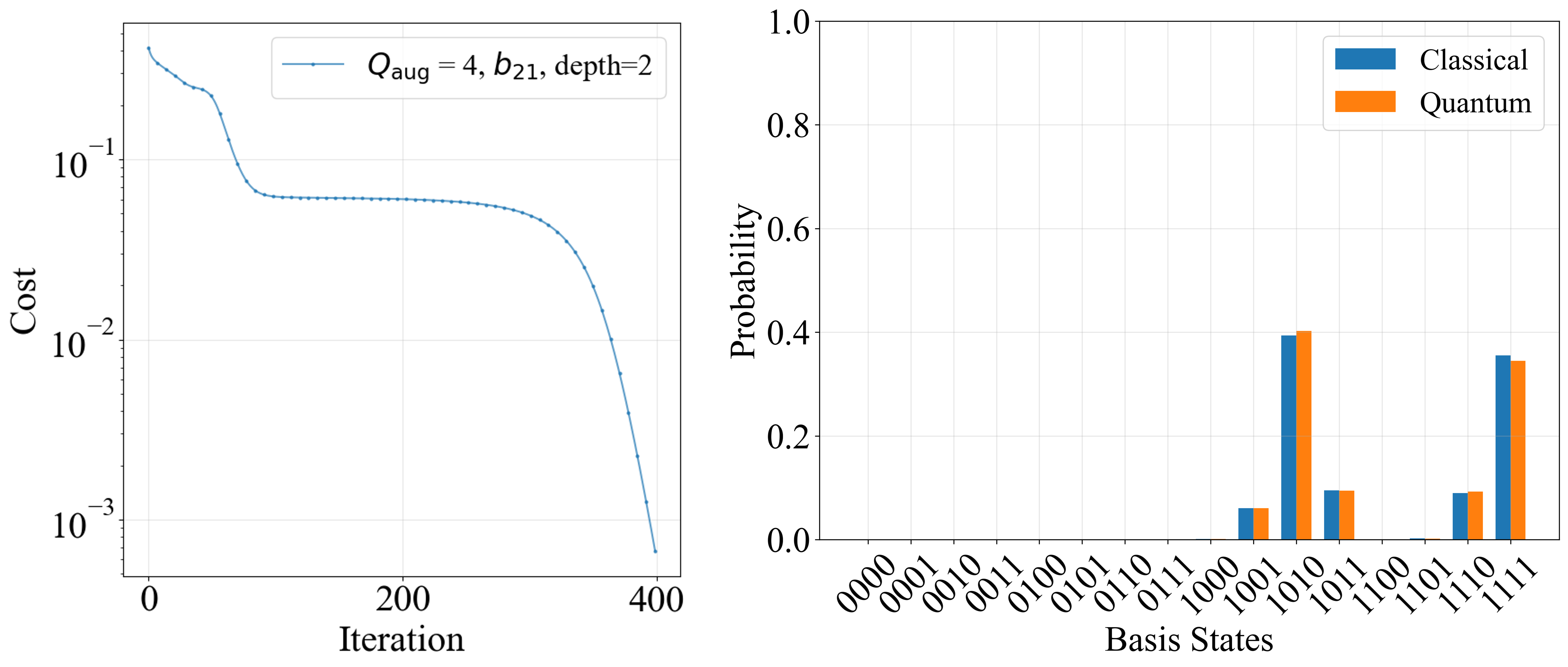}
    \caption{(a) Cost history and (b) heatmap of the highlighted example in Table III of the main text.}
    \label{fig:pl_b_Q4}
\end{figure}
